\documentstyle[prb,aps,twocolumn,psfig]{revtex}
\begin{document}
\wideabs{
\draft
\title{Two-dimensional electron system in high magnetic fields: 
Wigner crystal vs. composite-fermion liquid}
\author{Sudhansu S. Mandal$^{\dagger}$, Michael R. Peterson, and Jainendra K. Jain}
\address{Department of Physics, 104 Davey Laboratory,
The Pennsylvania State University, University Park, Pennsylvania 16802}
\address{$^{\dagger}$ Theoretical Physics Department, Indian Association
for the Cultivation of Science, Jadavpur, Kolkata 700 032, India}
\date{\today}
\maketitle

\begin{abstract}
The two dimensional system of electrons in a high magnetic field offers an opportunity
to investigate a phase transition from a quantum liquid into a Wigner solid. 
Recent experiments have revealed an incipient composite fermion liquid in a 
parameter range where theory and many experiments had previously suggested 
the Wigner crystal phase, thus calling into question our current understanding. 
This Letter shows how very small quantitative corrections ($<1$\%)
in the energy due to the weak interaction between composite fermions
can cause a fundamental change in the nature of the ground state,
thus providing insight into the puzzling experimental results.
\end{abstract}
}

When electrons confined in two dimensions are exposed to a strong magnetic field, 
they fall into the lowest quantum of their kinetic energy (Landau level), 
and the physics is governed by the interaction alone.  With the kinetic energy 
thus quenched, one might have expected electrons immediately to form a lattice, known 
as the Wigner crystal \cite{Wigner}, but experiments discovered the fractional quantum Hall
effect \cite{Tsui} instead, which is understood in terms of a correlated liquid of 
composite fermions \cite{Jain,Pinczuk,Heinonen,Jain00}.  The phase 
transition from this exotic quantum liquid into the electron solid anticipated at 
yet higher magnetic fields has been a subject of intense investigation.

\begin{figure}
\centerline{\psfig{figure=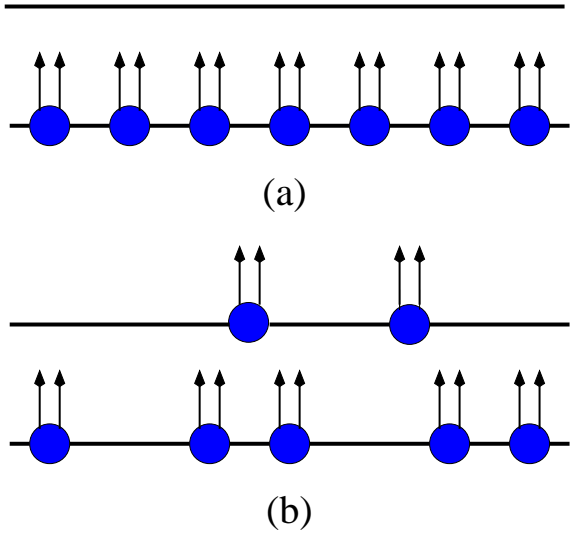,width=3.0in,angle=-90}}
\hspace{5cm}
\caption{
Examples of configurations of $^2$CFs at $\nu^*=1$
($\nu=1/3$).  Panel (a) shows pictorially the ground state of noninteracting composite fermions, 
while panel (b) shows two $^2$CF particle-hole pairs.  The analogous configurations at 
$\nu=1/7$, 1/9, and 1/11 will contain $^6$CFs, $^8$CFs, and $^{10}$CFs, respectively.
\label{jain_fig1}}
\end{figure}

In a range of filling factor $\nu$, defined as the ratio of the number of 
particles to the Landau level (LL) degeneracy, 
interacting electrons in the lowest LL capture an even number of quantized vortices to 
transform into composite fermions (CFs), because that is how they avoid one 
another most effectively.  There are many types of composite fermions distinguished
by their vorticity $2p$, denoted by $^{2p}$CFs.  The effective filling factor of $^{2p}$CFs, 
$\nu^*$, is given by the relation $\nu=\nu^*/(2p\nu^*\pm 1)$.  (The CFs are often thought of  
as electrons carrying an even number of magnetic flux quanta; see Fig.~\ref{jain_fig1}).
It is often a good first approximation to treat the 
composite fermions as noninteracting.
The experimental observation of fractional quantum Hall effect (FQHE)
at $\nu=n/(2pn\pm 1)$ is well explained as the 
$\nu^*=n$ integral quantum Hall effect \cite{Klitzing} of $^{2p}$CFs, 
and extensive studies \cite{Pinczuk,Heinonen,Jain00} have 
shown that the model of noninteracting composite fermions is also quite 
accurate quantitatively.  In particular, Laughlin's trial wave function \cite{Laughlin} 
at $\nu=1/m$, where $m=2p+1$ is an odd integer, is interpreted as the ground state 
of noninteracting $^{2p}$CFs at $\nu^*=1$.  Variational studies comparing its energy  
with the energy of the Wigner crystal have found \cite{Lam,Chui1,Zhu,Yang,Fertig}
that the liquid wins for 
$\nu=1/3$ and 1/5, but the WC has lower energy at $\nu=1/7$ and smaller filling factors.  
Although the WC has so far eluded a {\em direct} observation, 
the insulating state observed in the region $\nu<1/5$ has been  interpreted as a 
Wigner crystal pinned by disorder \cite{Jiang,GoldmanWC,Shayegan}.  
However, transport \cite{Goldman17} and 
photoluminescence \cite{Kukushkin} experiments
reported evidence for FQHE-like structure in the neighborhood of $\nu=1/7$ and 
$\nu=1/9$ more than a decade ago. The recent observation \cite{Pan}, 
in exceptionally high quality samples, of a series of resistance minima
at filling factors $\nu=1/7$, 2/11, 2/13, 3/17, 3/19, 1/9, 2/15  and 2/17 
has given a particularly compelling indication for the existence of an incipient 
liquid of $^6$CFs and $^8$CFs in the range $1/5>\nu \geq 1/9$.

While the approximation of neglecting the weak residual interaction between 
the composite fermions is accurate, it is not exact.  Because 
even small quantitative discrepancies can have  
substantial qualitative consequences near a phase boundary, we proceed below to  
calculate corrections due to the inter-CF interaction to determine how it modifies the 
previous results.

The calculations will be performed in the spherical geometry, where 
$N$ electrons are confined to the two-dimensional surface of a
sphere, moving under the influence of a radial magnetic field produced by 
a magnetic monopole of strength $Q$ at the center.  According to Dirac's quantization 
condition, $Q$ can be either an integer or a half integer,
and produces a total flux of $2Q\phi_0$, where $\phi_0=hc/e$ is called the flux quantum.
Interacting electrons at $Q$ map into  weakly interacting $^{2p}$CFs at $Q^*=Q-p(N-1)$.  
We will investigate the $n=1$ state of
$^6$CFs, $^8$CFs, and $^{10}$CFs, corresponding to FQHE at
$\nu=1/7$, 1/9, and 1/11, respectively.

The system of noninteracting composite fermions is {\em defined} by analogy to 
noninteracting electrons at $\nu^*$.  In particular, the 
ground state of noninteracting composite fermions at $\nu^*=n$ is 
obtained from the ground state of noninteracting electrons at $\nu^*=n$, 
namely the state with $n$ filled Landau levels, $\Phi_n$, as 
\begin{equation}
\Psi^{(0)}=P_{LLL}\Phi_1^{2p} \Phi_n
\end{equation} 
where $\Phi_1$ is the wave function of the lowest fully occupied Landau level 
and $P_{LLL}$ is the lowest LL projection operator.  The factor 
$\Phi_1^{2p}$ attaches $2p$ vortices to each electron to convert it into a 
composite fermion, so the wave function on the right hand side may be viewed 
as $n$ filled levels of composite fermions.  
For $n=1$, the above wave function, shown pictorially in Fig.~1a, 
reduces to Laughlin's wave function.  The residual interaction between 
composite fermions modifies $\Psi^{(0)}$ through hybridization with higher CF levels.  
We consider 
\begin{equation}
\Psi_{\alpha}^{(j)}=P_{LLL}\Phi_1^{2p} \Phi_{\alpha}^{(j)}
\label{eq2}
\end{equation}
where $\Phi_{\alpha}^{(j)}$ denote states obtained from $\Phi_n$ by exciting $j$ electrons
from the topmost occupied LL  
to the lowest unoccupied LL, with $\alpha$ labeling distinct particle-hole 
configurations.  $\Psi_{\alpha}^{(j)}$ thus contain $j$ 
particle-hole pairs of composite 
fermions (Fig.~\ref{jain_fig1}).  An improved representation for the 
ground state, $\chi^{(J)}$, which involves the effect of CF level mixing caused by the 
inter-CF interaction, 
is obtained by diagonalizing the Coulomb Hamiltonian in the basis given by 
$[\Psi^{(0)}, \{\Psi_{\alpha}^{(1)}\},  \{\Psi_{\beta}^{(2)}\}, ...
\{\Psi_{\gamma}^{(J)}\}]$.  The 
corresponding energy per particle, $E^{(J)}$, gives a strict variational upper 
bound for the true ground state energy.  
Successively higher values of $J$ produce lower energies, and convergence 
is achieved when $E^{(J)}$ becomes more or less independent of $J$.

With explicit wave functions in hand, the 
computation is conceptually straightforward, but technically challenging due to the 
complex form of wave functions.  Several 
complications are handled as follows.  The lowest LL projection 
is evaluated as discussed in the literature \cite{JK}, by replacing the single 
particle eigenstates 
in $\Phi_{\alpha}^{(j)}$ by appropriate wave functions for  
composite fermions.  In the spherical geometry, the liquid state, which possesses 
rotational symmetry, has total orbital angular momentum $L=0$, as appropriate for a filled 
angular momentum shell.  The Coulomb interaction does not mix states
with different $L$, so it is convenient to work within the $L=0$ sector. 
We generate the many body states $\Phi_{\alpha}^{(j)}$ in this sector by 
diagonalizing the Coulomb interaction at $\nu^*=n$ in the subspace containing 
$j$ particle-hole pairs.  The resulting eigenstates are a linear superposition of Slater 
determinant basis states, from which basis functions at $\nu$ are obtained 
according to Eq.~(\ref{eq2}).

The angular momentum of each successive LL increases by one unit in 
the spherical geometry.  For the states with a single particle hole pair, the 
angular momenta of the particle and the hole differ 
by one unit, and as a result fail to produce a state with total angular momentum $L=0$.
It is therefore necessary to consider  
$J\geq 2$.  We consider in this work up to three particle hole pairs ($J=3$); 
for still higher values of $J$, 
the basis becomes too large to produce numerically stable energies.  
Our inability to go beyond $J=3$ imposes a limitation on the system size for which we 
can obtain reliable results.

Another major obstacle is that the basis functions at $Q$ 
generated from the wave functions at $Q^*$ are not orthogonal.  The standard Gram-Schmidt 
procedure will be employed to construct an orthogonal basis.  The inner products needed 
for this purpose, as well as the matrix elements of the Coulomb
Hamiltonian needed for diagonalization are computed by Monte Carlo. 
Approximately 10 Monte Carlo runs were performed for each energy, with 0.5-1.0 
$\times 10^6$ iterations in each run.  The calculation was performed on a PC cluster, 
and took the equivalent of $>$ 100 days of CPU time on a single node (dual 1 GHz 
Intel PIII). The statistical uncertainty in each matrix element translates into an uncertainty 
in the energy, which we
estimate by calculating it in $\sim$ 10 different runs.  Satisfactory accuracy is obtained with the 
number of iterations performed.  More details on the method can be found elsewhere \cite{Mandal}. 
All energies quoted below include interaction with a uniform, positively charged background, and 
have been multiplied by a factor $\sqrt{2Q\nu/N}$ to correct for the finite 
size deviation in the density.  A strictly two dimensional system is assumed 
and a mixing between the LLs of {\em electrons} is neglected.

There are substantial fluctuations in energy for $N\leq 6$,
which we believe arise from commensurability effects.
For $N=4$ and 6, polyhedral arrangements are possible on the sphere (tetrahedran and octahedran) 
producing anomalously low energies.  For $N=3$ there is only a single $L=0$ state at $\nu=1/m$, 
indicating that the system is too small to be meaningful.  In the following, we only consider 
$N\geq 7$, where we have found that the ground state energy behaves more or less smoothly 
with $1/N$.  (Polyhedra are possible also for certain larger values of $N$, but 
there the quantitative effect is small.)

\begin{figure}
\centerline{\psfig{figure=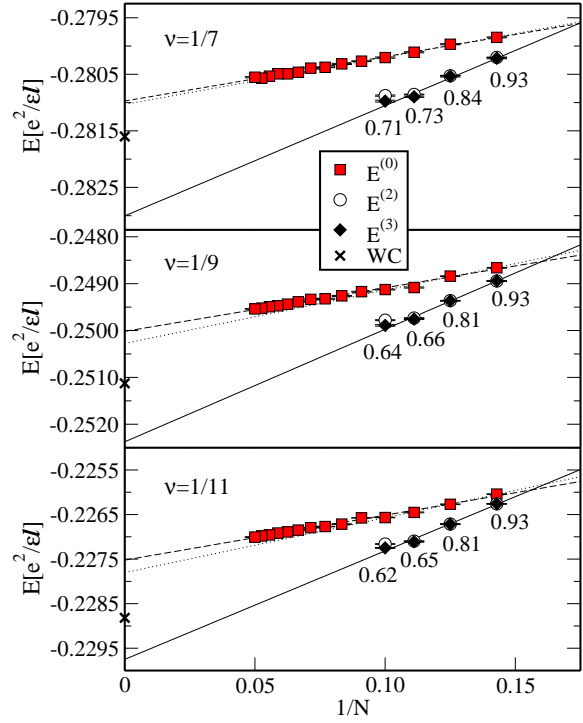,width=3.5in,angle=0}}
\caption{The ground state energy per particle   
as a function of the number of composite fermions, $N$, for $\nu=1/7, 1/9,$ and $1/11$.  
$E^{(0)}$ is the energy of the noninteracting CF system, while $E^{(2)}$ and $E^{(3)}$ are 
successively better approximations including interaction effects.  
The cross on the y-axis shows the energy 
of the Wigner crystal, taken from Ref.~\protect\onlinecite{Lam}, which has an 
uncertainty of $\sim$ 0.0005 in the chosen units.
The error bars denote the statistical uncertainty in the Monte Carlo 
and the solid straight lines are the best chi-square fits for $E^{(3)}$.
The number near each solid diamond equals
$|<\chi^{(3)}|\Psi^{(0)}>|^2/(<\chi^{(3)}|\chi^{(3)}><\Psi^{(0)}|\Psi^{(0)}>)$.
\label{jain_fig2}}
\end{figure}

We begin by testing the accuracy of the method.
Table 1 shows $E^{(0)}$ and $E^{(2)}$ for several $N$ at 
$\nu=1/5$, along with exact energies obtained by 
numerical diagonalization.  $E^{(2)}$ has a remarkable accuracy of 
0.01\% or better, which is a significant improvement over $E^{(0)}$, which 
is off by up to 0.2\%.   
Fig.~(\ref{jain_fig2}) shows the energies $E^{(0)}$,  $E^{(2)}$, and $E^{(3)}$ as 
a function of $N^{-1}$ for $\nu=1/7$, 1/9, and 1/11.  For $N=7$, 8, and 9, 
convergence is manifest.  For $N=10$, $E^{(J)}$ is not yet $J$ independent, but we expect 
$E^{(3)}$ to be reasonably accurate. 
Larger values of $J$ must be considered with increasing $N$ in order to obtain meaningful
estimates;  for a fixed $J$, the difference between the energies (per particle) 
$E^{(0)}$ and $E^{(J)}$ is O($N^{-1}$), which vanishes in the thermodynamic limit. 
With $J\leq 3$, we are able to obtain reliable estimates for the energy for only up to $N=10$.
(It is noted that the dimension of the full Fock space for 10 particles at $\nu=1/11$ 
is $1.6 \times 10^{13}$, rendering exact diagonalization impracticable.)

We use a linear extrapolation in $N^{-1}$ to obtain the
thermodynamic limit (solid line), as appropriate for a compact geometry without 
boundaries (boundaries 
would contribute a term $\sim N^{-1/2}$ in the finite size correction for the energy per 
particle).  To get a feel for the accuracy of such an extrapolation with four 
points, we note that for $E^{(0)}$, which serves as a useful benchmark,
the linear extrapolation using the first four points (dotted line) 
differs from the more accurate extrapolation (dashed line) by approximately 
0.04\%, 0.1\%, and 0.15\% for $\nu=$ 1/7, 1/9, and 1/11, respectively.

$E^{(0)}$, the energy of Laughlin's wave function, exceeds the WC energy 
for $\nu=1/7$, 1/9, and 1/11,
as noted in earlier studies \cite{Lam,Chui1}.  However, the best variational estimate,  
$E^{(3)}$, is $\sim$ 0.8\% lower than $E^{(0)}$
for the filling factors considered, which is not a large reduction in absolute terms, 
but enough to make a qualitative change in the outcome.
The principal conclusion of this work is that, for the model considered, the 
ground state is a liquid at $\nu=1/7$ and very likely also at $\nu=1/9$.
At $\nu=1/11$ the relatively large uncertainties in various energies prevent us from making a 
definite statement, but a liquid ground state is a distinct possibility even here.  
It is striking that composite fermions with vorticity as high as eight or perhaps even ten 
can occur in nature.

The relatively small overlaps between $\Psi^{(0)}$ and $\chi^{(3)}$
(Fig.~\ref{jain_fig2}) are indicative of a substantial amount of mixing between CF levels.
Nevertheless,   there is little doubt that the true ground state can be obtained 
from the noninteracting state $\Psi^{(0)}$ without crossing a phase boundary (or without
gap closure), and hence the physics of the true liquid state can {\em qualitatively} be 
described in terms of noninteracting composite fermions.
Only detailed and accurate quantitative calculations can tell, however,
whether the liquid or the solid is the ground state.

The local resistance minima observed at fractions of the form  
$\nu=n/(6n\pm 1)$ and $\nu=n/(8n\pm 1)$ ride a 
background that rises rapidly with decreasing temperature \cite{Pan}.  
Pan {\em et al.} \cite{Pan} suggested that the CF liquid is  
obtained at temperatures exceeding the WC melting temperature.  Our calculations  
put forth the alternative possibility that the true ground state at these fractions,
in the absence of disorder, is the CF liquid.  The insulating behavior would then be the result  
of an absence of percolation of the liquid, due to domains of Wigner crystal 
induced by disorder \cite{Chui}.  A future observation of a well developed, 
dissipationless FQHE state at these fractions would confirm this scenario.
(It may be recalled that the FQHE at $\nu=1/5$ was also initially observed on a rising background.)
Interesting questions also arise regarding the nature of the state {\em away} from these 
special filling factors. 
It is possible that there is a series of zero-temperature quantum phase transitions between
the CF liquid and the WC as a function of $\nu$, as is 
believed to be the case \cite{Jiang} in the vicinity of $\nu=1/5$.

To summarize, the interaction between composite fermions is weak, 
causing only a small change in the energy ($<1$\%), but it can be of crucial 
importance in stabilizing the CF liquid in parts of 
the phase diagram earlier thought to belong to the Wigner crystal. 
We thank Vito Scarola for help and useful discussions.
This work was supported in part by the National Science Foundation under grants
no. DGE-9987589 (IGERT) and DMR-0240458.  We are grateful to the High Performance
Computing (HPC) Group led by V. Agarwala, J. Holmes, and J. Nucciarone, at the 
Penn State University ASET (Academic Services and Emerging Technologies) for
assistance and computing time with the LION-XE cluster.

\begin{table}[t]
\caption{The ground state energy per particle for $N$ electrons at $\nu=1/5$.  
The Laughlin energy $E^{(0)}$ (also the energy of the noninteracting CF ground
state), and $E^{(2)}$, which includes the effect of inter-CF interaction at the lowest level
of approximation, 
are compared with the exact energy obtained from numerical diagonalization. 
The energies are quoted in units of $e^2/\epsilon l$ ($l\equiv\sqrt{\hbar c/eB}$ is the magnetic
length and $\epsilon$ is the dielectric constant of the host semiconductor), and the 
statistical uncertainty from the Monte Carlo simulation is shown in parentheses.
For three particles, $E^{(0)}$ and $E^{(2)}$ are exact for the trivial reason that  
there is only one uniform state.
\label{tab1}}
\begin{center}
\begin{tabular}{|c|c|c|c|} 
$ N$ & $E^{(0)}$ &  $E^{(2)}$        &  $E^{exact}$  \\ \hline
3 & -0.325428(14)&   -0.325422(14) & -0.325431  \\ \hline
4 & -0.326531(15) &   -0.326965(10) &  -0.326992  \\ \hline
5 & -0.326490(17)&   -0.326577(14) & -0.326586  \\ \hline
6 & -0.326805(18)&   -0.327424(14) &    -0.327462 \\
\end{tabular}
\end{center}
\end{table}

\pagebreak

\end{document}